\documentclass[12pt]{article}
\usepackage{amsmath}
\usepackage{amssymb}
\usepackage{epsfig}
\usepackage{cite}
\usepackage{color,colordvi}

\begin{document}
\vspace{0.01cm}
\vskip-0.5cm
\begin{center}
{\Large\bf  Non-Thermal Corrections to Hawking Radiation Versus the Information Paradox\footnote{Based on discussion 
given at 2015 Erice Summer School "Future of Our Physics Including New Frontiers."}}

\end{center}

\vspace{-0.1cm}
%\end{center}

\begin{center}

{\bf Gia Dvali}$^{a,b,c}$,

\vspace{.6truecm}

%\vspace{.2truecm}

{\em $^a$Arnold Sommerfeld Center for Theoretical Physics\\
Department f\"ur Physik, Ludwig-Maximilians-Universit\"at M\"unchen\\
Theresienstr.~37, 80333 M\"unchen, Germany}

%\vspace{.2truecm}

{\em $^b$Max-Planck-Institut f\"ur Physik\\
F\"ohringer Ring 6, 80805 M\"unchen, Germany}

{\em $^c$Center for Cosmology and Particle Physics\\
Department of Physics, New York University, \\ 
4 Washington Place, New York, NY 10003, USA}

\end{center}

\vspace{0.1cm}

\begin{abstract}
\noindent  
 {
\small
 We provide a model-independent  argument indicating that for a black hole of entropy $N$ the non-thermal deviations from Hawking radiation, per each emission  
 time, are of order $1/N$, as opposed to $e^{-N}$. This fact abolishes the standard {\it a priory}  basis for the information paradox. }

\end{abstract}

\thispagestyle{empty}
\clearpage

 The famous Hawking's information paradox\cite{paradox}  is based on an implicit assumption that 
 during the black hole evaporation process the deviations from  exact thermality  are negligible.
 On the other hand, this assumption does not hold in a microscopic quantum theory of a black hole, where the corrections are suppressed by inverse powers of entropy \cite{usBH}.  
 In this note, we shall distill the model-independent aspect 
of these corrections.  Without relying on any particular microscopic picture, 
 we shall show that the exact thermality assumption contradicts  to some 
  well-established knowledge about black holes and quantum physics. 
 The presented argument reflects on the discussions presented in series of papers written together with Cesar Gomez, such as\cite{usBH} 
 and subsequent articles.

  First, we shall recall that Bekenstein's entropy scales as area in Planck units \cite{Bek}, 
    \begin{equation}
 N \, =\, {R^2\over L_P^2} \, .
 \label{N}
 \end{equation}
  Here $R = G_NM$ is the gravitational radius of a black hole of mass $M$ and $G_N$ is the Newton's constant. 
  $L_P$ is the Planck length, which in terms of $G_N$ and Planck's constant $\hbar$ is defined as 
     \begin{equation}
 L_P^2 \, \equiv \, \hbar G_N \, .
 \label{LP}
 \end{equation}
  We set the speed of light and all the irrelevant numerical factors equal to one.  
  
   Next, let us define a rigorous limit, in which the assumption of exact thermality is correct. 
  This is the limit in which any quantum back-reaction to the black hole geometry can be ignored.
   For finite values of $R$ and $\hbar$, the only possible way to achieve this is to take the 
  double-scaling limit:   
 \begin{equation}  
  M \rightarrow \infty, ~ G_N \rightarrow 0, {\rm but~ keep }~ R ~{\rm finite} \, . 
  \label{dpuble}
 \end{equation}
 In this case, the back-reaction vanishes and the black hole radiates 
 in an exactly thermal spectrum of Hawking temperature \cite{hawking},  
 \begin{equation}
   T \, = \, {\hbar \over R} \, .
   \label{temperature} 
  \end{equation} 
 On average,  the black hole radiates one quantum per emission time $t_{emission} \sim R$.  
    
     Thus, the only situation in which the black hole radiation is exactly thermal, is when its entropy is infinite. 
    We therefore can use the quantity $N$ for parameterizing the  corrections to thermality for a black hole 
    of finite mass and entropy. 
    The crucial question is:  Are the corrections of ${1 \over N}$- or of $e^{-N}$-type? 
     As it will become evident, the former is true. 
   
    The argument consists of two parts. 
    First, let us estimate the time-variation of temperature. 
 Using the Stefan-Boltzmann law  for Hawking radiation, 
   \begin{equation}
     \dot{M} \, = \, -{1 \over \hbar} T^2 \, ,  
    \label{SB}
    \end{equation}
    and the relation (\ref{temperature}) between the temperature and the mass in the constant mass approximation,  
   we can easily derive the leading order effect,  
  \begin{equation}
   \hbar \, {\dot{T} \over  T^2} \, = \, {1\over N}  \,. 
  \label{change} 
  \end{equation}
  The quantity ${\hbar \dot{T} \over  T^2}$ is very important, because it tells us how rapid is the relative change of temperature.   For a generic thermal system this parameter measures the departure from the thermal equilibrium.  Usually, the system is unable to  track the change of temperature infinitely-closely and this results in non-thermal corrections.  As we see, in case of a black hole, this parameter is not exponentially small, but is only $1/N$-suppressed.  This should already raise a flag about the validity of the exact thermal approximation. 
   However, to complete our argument we have to exclude the following loophole.  
  
  Namely, one may argue that the internal qubit degrees of freedom, which store black hole information and 
  reveal it through the deviations from Hawking's thermal spectrum, are able to thermalize quickly-enough in order to  track the varying temperature exponentially-closely.  
  
   The indication that this is impossible follows from the scaling of black hole entropy in $\hbar \rightarrow 0$ limit. 
   As it is clear from (\ref{N}), in this limit the entropy becomes infinite. The only way to reconcile this with the classical no-hair properties is to 
   accept that the information-storing degrees of freedom interact with a $1/N$-suppressed strength.  In such a case 
   the information-retrieval time  becomes infinite when $N \rightarrow \infty$ and the classical black hole hair becomes undetectable.  
    But, then it is impossible to reconcile such a weak interaction strength with the possibility 
    of exponentially-fast thermalization\footnote{This is fully supported by the explicit microscopic 
    model \cite{usBH}  in which the chaos and thermalization-type time-scales, such as, the scrambling 
    time  $t_{scr} = R ln(N)$  \cite{scrambling} or the maximal-entanglement time \cite{Page}, become infinite for  $N=\infty$.}. 
     Under any sensible estimate, the characteristic qubit-qubit interaction time  is at least $t_{int} \sim R N^{2}~$.\footnote{Notice, that as a consistency check, this interaction rate gives 
  a correct Hawking emission time, because the time-scale during which at least one pair of qubits will
  interact is obtained via dividing $t_{int}$  by number of pairs ($n_{pairs} \sim N^2$), which gives precisely the Hawking emission time, $t_{emission}  = t_{int} /n_{pairs} = R$ \cite{usBH}.}  Obviously, with such a slow interaction, the full thermalization time is way longer  than  the black hole emission time $t_{emission} \, = \,  R$ . 
 For example, already the scrambling time, is much longer,  $t_{scr} = R{\rm ln} (N)$ \cite{scrambling}. 
  Thus, 
     it is highly unlikely that qubit degrees of freedom could re-thermilize over such a short time-period. 
     Moreover, such a fast re-thermalization would violate unitarity bound on fast scrambling\footnote{Indeed, 
  if the scrambling time were equal to $t_{emission}$, this would imply that it is not growing with the number of qubits, $N$, which would mean that their coupling increases 
  with $N$ and becomes strong.}.           
Hence, the non-thermal corrections of order (\ref{change})  are inevitable.     
     This completes our argument. 
     
     To summarize, we argued that in a sensible quantum theory the internal black hole degrees of freedom are unable to maintain the exact thermality of Hawking radiation and simultaneously  accommodate the correct scaling of entropy. 
  Resolution of the conflict is in accepting $\sim {1 \over N}$ non-thermal deviations from Hawking radiation 
  that carry black hole hair.                 
      These deviations, by magnitude, 
     are fully sufficient for encoding the entire black hole information, because the total number of emission acts 
    is of order $N$ (e.g., see the last ref. in \cite{usBH} for counting).  It is not the goal of the present note to discover how these corrections encode information. 
     For this, one has to employ a specific microscopic theory, such as, e.g., \cite{usBH}.   
        
    Our message is that the resolution of the information paradox must lie in taking into account the large ${1 \over N}$-deviations from Hawking spectrum, 
    rather than in more dramatic possibilities, such as, giving up the unitarity.

\section*{Acknowledgements}
 
This note is based on the discussion organized by Prof. Antonino Zichichi at Erice summer school
"Future of Our Physics Including New Frontiers". It is a real pleasure to thank him together with Prof. Gerard 't Hooft for an interesting discussion.  
 We thank Cesar Gomez for many valuable discussions and collaboration.   
This work was supported by Humboldt Foundation under Alexander von Humboldt Professorship,  by European Commission  under ERC Advanced Grant 339169 ``Selfcompletion'' and  by TRR 33 "The Dark
Universe".

\end{document}